\documentclass[epj]{svjour}
\usepackage{graphics}
\begin{document}
\title{ Localization, Dirac Fermions and Onsager Universality}
\author{Indubala I. Satija\cite{www}}
\institute{
 Department of Physics, George Mason University,
 Fairfax, VA 22030}
\date{Received: date / Revised version: date}
\abstract{
Disordered systems exhibiting exponential localization are mapped to anisotropic spin chains
with localization length being related to the anisotropy of the spin model.
This relates localization phenomenon in fermions to the rotational symmetry breaking
in the critical spin chains. 
One of the intriguing consequence is that the statement of Onsager universality 
in spin chains implies universality of the localized fermions where the fluctuations
in localized wave functions are universal. 
We further show that the fluctuations about localized nonrelativistic fermions describe relativistic fermions.
This provides a new approach to understand the absence of localization in
disordered Dirac fermions.
We investigate how disorder affects well known universality
of the spin chains by examining the multifractal exponents. Finally, we examine the effects
of correlations on the localization characteristics of relativistic fermions. 
\PACS{
      {PACS-key}{Harper equation, localization}   \and
      {PACS-key}{Dirac fermions}
     } 
} 
\authorrunning
\titlerunning
\maketitle
Two-dimensional Ising model is one of the few examples of exactly solvable many body systems.\cite{book}
The model exhibits a phase transition at finite temperature characterized by universal exponents defining
a universality class, the Onsager universality which describes the phase transitions for anisotropic XY models.
Interestingly, the Onsager universality
also describes a quantum phase transition at zero temperature, driven by quantum fluctuations, 
of one-dimensional quantum anisotropic XY spin chains in a transverse field.\cite{Lieb} 
These quantum models belong to a small family of integrable Hamiltonians that have attracted both
theoreticians as well as experimentalists.
Heart of integrability of these many body quantum spin problem is a mapping between spin and fermions
which relates interacting XY spin chain with $O(1)$ symmetry  to the fermion Hamiltonian that
are quadratic in fermions. The
spin-fermion correspondence has proven to be extremely important also for the case of disordered magnetic field
as the quadratic fermion Hamiltonian can be numerically diagonalized with extreme precision.\cite{IIS}
Recent studies have shown that the a large variety of disordered quantum chains with $O(1)$ spin symmetry
are still described by the Onsager
universality.\cite{Luck,IIS}

In this paper, we show a new type of spin-fermion mapping where
disordered fermions exhibiting
exponential localization are shown to be related to anisotropic spin chain
in a disordered magnetic field at the onset to long range order(LRO). 
This correspondence, valid exclusively for exponentially localized systems,
provides new insight into various issues relevant
to disordered fermion as well as spin models. In particular we exploit well established universality
hypothesis of spin systems to make important statements about fermion problems.
Firstly, the root of recently observed universality in localized fermions\cite{KSprl,andy}
is traced to the Onsager universality of the spin systems.
Another interesting result is a correspondence between the relativistic and the
nonrelativistic fermions in the presence of disorder.  
It is shown that the relativistic fermions can be viewed as
the fluctuations in the exponentially localized solutions of the
nonrelativistic fermions. This provides a new approach for understanding
the absence of localization in disordered Dirac fermions
which has been the subject of various recent studies.\cite{Dirac} 
We argue that the long range magnetic correlations provide mechanism for delocalization of relativistic fermions
thus obtaining a deeper understanding of the absence of localization.
In addition to obtaining intuitively appealing picture
of some surprising results of disordered fermions, we also obtain a generalization of universality
statement of the critical exponents for disordered spin models.
Finally, we examine how the correlations affect the localization characteristics
and show the possibility of delocalization of relativistic fermions analogous to the corresponding
nonrelativistic case.

Although the setting we describe is quite general in the context of disordered systems,
for concreteness we will consider
quasiperiodic disorder where the lattice problem for nonrelativistic fermion is described by the
Harper equation,\cite{Harper}
\begin{equation}
\psi_{n-1}+\psi_{n+1}+2 \lambda V_n \psi_n = E \psi_n.
\end{equation}
Here $V_n = cos(\theta_n)$ where $\theta_n= (2 \pi \sigma_g n +\phi)$. The $\sigma_g$ is an irrational number
describing competing length scales in the problem and $\phi$ is a constant phase factor.
Harper equation, in one-band approximation describes
Bloch electrons in a magnetic field. This problem frequently arises in many different physical contexts,
every time emerging with a new face to describe another physical application.
The problem is solvable by Bethe-ansatz\cite{BA}: there is an algebraic Bethe-ansatz equation for the spectrum.
It was shown that at some special points in the spectrum, e.g at the mid-band points,
the Hamiltonian in certain gauge can be written as a linear combination of generators of a
quantum group called $U_q(sl_2)$. It also describes
some properties of the integer quantum Hall effect: it showed that Kubo-Greenwood
formula for the conductance of any filled isolated band is an expression for a topological
invariant, and is an integer multiple of $e^2/h$.
It was further shown by Avron et al\cite{Chern} that it is a topological invariant that defines the first
Chern class of the mapping of the Brillouin zone( a two-dimensional torus)
onto a complex projective space of the wave functions.

In contrast to usual Anderson problem describing localized particle in a random potential, the Harper equation
exhibits localization-delocalization transition provided $\sigma_g$ is an irrational number
with good Diophantine properties (ie. badly approximated by rational numbers).
This transition at $\lambda=1$ has been characterized with
singular continuous states and spectra. Richness and complexity of the critical point
describing localization transition has been studied in great detail by various renormalization
group approaches.\cite{Ostlund,KSrg} Recently, it was shown that multifractal characteristics
continue to exist beyond the critical point\cite{KSprl} throughout the localized phase.
This hidden complexity of the localized phase is brought to light after one factors out the
exponentially decaying envelope.
The localized wave functions with inverse
localization length $\xi^{-1}=log(\lambda)$ is rewritten as,\cite{KSprl}
\begin{equation}
\psi_n = e^{ -\gamma |n|} \eta_n
\end{equation}
where $\gamma=\xi^{-1}$.
The tight binding model(tbm) describing the fluctuations $\eta_n$ in the exponentially decaying
envelope is given by the following pair of equations,
\begin{eqnarray}
e^{-\gamma} \eta^r_{n+1} + e^{\gamma} \eta^r_{n-1} + 2 \lambda V_n
\eta^r_n &=& E \eta^r_n \nonumber\\
e^{\gamma} \eta^l_{n+1} + e^{-\gamma} \eta^l_{n-1} + 2 \lambda V_n
\eta^l_n &= &E \eta^l_n
\end{eqnarray}
Here $\eta^r$ and $\eta^l$ respectively describe the fluctuations 
to the right and to the left of the localization
center. An exact renormalization
scheme\cite{KSprl} showed that these fluctuations exhibit universal features(see Fig. 1)
described by the strong coupling fixed point. Hence, localized phase is characterized by
universal fractal characteristics described by $\lambda \to \infty$ limit of the equation.
These results were further confirmed by rigorous mathematical analysis.\cite{andy}
Hence the strength of the disorder which determines the localization length can be factored out making 
universal aspects transparent. 
\begin{figure}
\resizebox{0.5\textwidth}{!}{%
  \includegraphics{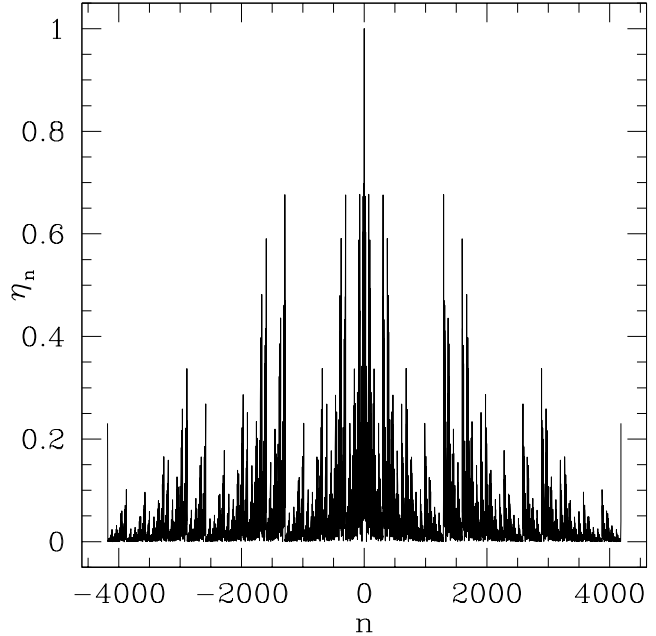}
}
\caption{Absolute value of the fluctuations for Harper equation for $E=0$ states with $\phi=.25$
and $\sigma_g=(\sqrt(5)-1)/2$. At the Fibonacci sites,
we see period-$6$ behavior(period-$3$ in absolute value): $|\eta_{F_m}|=|\eta_{F_{m+3}}|$ which
is independent of $\lambda$.}
\label{fig:1}
\end{figure}

A new revelation that provides intuitive understanding of
the strong coupling fixed point of Harper is obtained
by relating the fluctuations described by (3) to the anisotropic spin chain at the onset to LRO.
It turns out that the equations (3) for $E=0$ describe quasiparticle excitations of a 
$critical$ anisotropic XY spin-$1/2$ chain in a transverse magnetic field, given by the following spin
Hamiltonian.
\begin{equation} \label{spin}
H=-\sum [e^{-\gamma} \sigma^x_n \sigma^x_{n+1}+ e^{\gamma}\sigma^y_n \sigma^y_{n+1} + 2\lambda V_n \sigma^z_n]. 
\end{equation}
The $\sigma_k$, $k=x,y$ are the Pauli matrices.
The $e^{-\gamma}$ and $e^{\gamma}$ respectively describe the exchange interactions along the
$x$ and the $y$ directions in spin space.
Using Jordan-Wigner transformation, 
the interacting spin problem can be mapped to non-interacting
spinless fermion problem\cite{Lieb} where fermions are the quasiparticle
excitations of the spin chain obeying the following coupled equations,

\begin{eqnarray}\label{fermion}
e^{-\gamma} \eta^1_{l+1}+e^{\gamma}  \eta^1_{l-1}+2\lambda V_l \eta^1_l &=& E \eta^2_l \nonumber\\
e^{\gamma}  \eta^2_{l+1}+e^{-\gamma}  \eta^2_{l-1}+2\lambda V_l \eta^2_l &=& E \eta^1_l
\end{eqnarray}
At the onset to LRO, the excitation spectrum becomes gap-less and hence for the
$E=0$, the massless mode, the above two equations are degenerate
coinciding with the equation (3). Therefore, the massless excitations
of the critical anisotropic spin chain describe the
fluctuations in the exponentially localized excitations of the isotropic chain where the anisotropy
parameter is related to the localization length.

Therefore, the localization length like the anisotropy parameter is an irrelevant variable
and hence the statement of strong coupling universality of the Harper equation is synonymous with
the statement that anisotropic spin chain is in the universality class of the Onsager universality.
In short, we establish an equivalence between Ising fixed point of the anisotropic spin chain and the strong
coupling fixed point of the Harper equation
and thus provide a simple interpretation of the strong coupling universality of the Harper equation.
Furthermore, by relating anisotropy to the localization length, we obtain a new picture of the localized
phase: the role of the strength of the disorder is similar to that of the strength of the anisotropy,
which is an irrelevant parameter for renormalization flow. 
It should be noted that the spin-fermion mapping provides a new method to determine the localization length of
the tbms in the presence of disorder. 
The expression for the localization length, $\gamma=log(\lambda)$
can be viewed as the relation describing the
critical point of the spin chain in a
disordered field.

We next show that the fluctuation $\eta_n$ obey Dirac equation
for zero energy states.
In the long wave length limit, the equations(3) reduce to the Dirac equation.
We replace $n$ by $x$ and write
$\eta_{n\pm 1} = e^{\pm ip} \eta(x)$, where $p$ is the momentum canonically conjugate to $x$.
The equation (4) for the fluctuations for $E=0$ state can be described by the following non-Hermitian Hamiltonian $H_{fluc}$
and its adjoint,
$H_{fluc} = e^{-\gamma} e^{i p} + e^{\gamma} e^{-i p} + 2\lambda V(x)$.
In the limit ($p \rightarrow 0$),
the system for $E=0$ reduces to the Dirac equation,
\begin{equation}
[g \sigma_x p -i(2\lambda V(x)+2 cosh(\gamma) )\sigma_y]\eta(x) = 0
\end{equation}
where $\eta(x)$ is a two-dimensional spinor
$\eta(x)=(\eta^l(x),\eta^r(x))$.
It is interesting that the two-component structure of Dirac spinor arises naturally when we consider
fluctuations about exponentially
localized wave functions.
Here $g \equiv 2sinh(\gamma)$ is the velocity of the Dirac fermions while the mass of the Dirac fermions
$m(x) = 2((\lambda V(x)+ cosh(\gamma))$.
Therefore, on a lattice, the Dirac fermions
with disordered mass are the fluctuations of the nonrelativistic localized fermions.
This would imply the absence of exponential localization for relativistic fermions.
The defiance of localization by relativistic fermions has been the subject of various studies
and our analysis provides a simple way to understand this intriguing result.

Next, we address the question whether the strong coupling fixed point which describes localized
phase of Harper equation, the critical Ising model and the Dirac fermions with quasiperiodic disorder, implies universal
multifractal exponents.
We compute the $f(\alpha)$ curve( Fig.2) describing the multifractal spectrum
associated with the self-similar wave function or
the inverse participation ratios ${\bf P}$, 
${\bf P}(q,N) = \frac{\sum |\eta_n|^{2q}}{\sum |\eta_n|^2} \sim N^{-\tau(q)}$,
$\alpha = \frac{d\tau}{dq}$ and
$f(\alpha) = \alpha q-\tau(q)$
The free energy function $\tau(q)$ and
its Legendre transform $f(\alpha)$ were found to be $\lambda$ independent  only for {\it for positive values of $q$}
and hence only left half of the $f(\alpha)$ curve is universal.
Therefore, for quasiperiodic spin chains at the onset to LRO, scaling exponents for {\it only
positive moments} of the participation ratio are universal.
This can be viewed as a generalization of the universality statements of periodic spin chains
to disordered spins.
\begin{figure}
\resizebox{0.5\textwidth}{!}{%
  \includegraphics{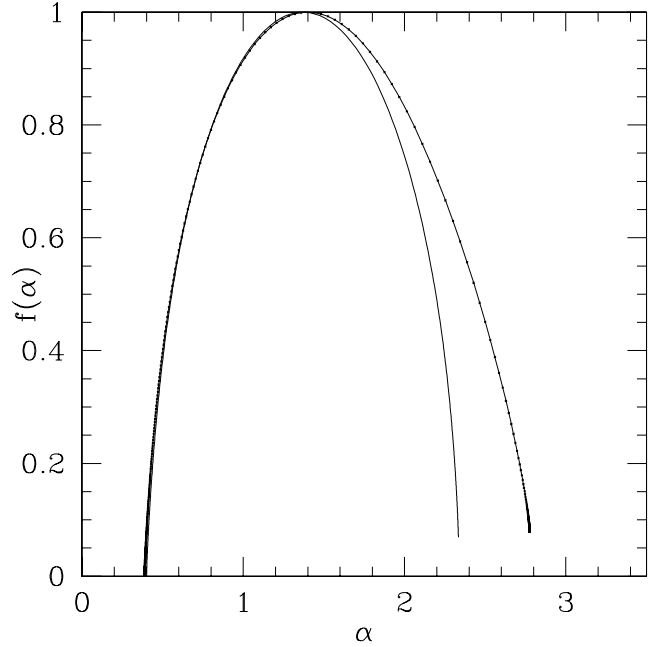}
}
\caption{Numerically obtained $f(\alpha)$ curves for $\lambda \rightarrow \infty$( solid curve) and
$\lambda=1.5$ (lines with crosses).}
\label{fig:2}
\end{figure}

Finally we investigate the role on correlations on the localization characteristics of massless
spin excitations which obey Dirac equation. The fact that correlations would result in delocalization
as originally shown by a random-dimer model\cite{pp}
is an important result in localization theory.\cite{pp}
For quasiperiodic disorder, dimer-type correlations can be introduced by
replacing $\theta_n=2\pi\sigma_g n$ in $V(\theta_n)$ by the iterates of the $supercritical$
standard map, describing
Hamiltonian systems with two degrees of freedom.\cite{dimerlong}
\begin{equation}
\theta_{n+1}+\theta_{n-1}-2\theta_n = - {K\over  2\pi } \sin( 2 \pi \theta_n) .\label{SM}
\end{equation}
We use iterates that describe golden-mean cantorus(the remanent of the KAM torus beyond the onset to global stochasticity)
which has been shown to exhibit dimer-type correlations, and leads to
Bloch-type states for the nonrelativistic fermions.\cite{dimerlong}
Here, we will confine to the Ising limit( can be obtained from (5) using $\gamma \to \infty$ and rescaling the parameters), described by,
\begin{eqnarray}
\eta^1_{n+1}+2\lambda cos(2\pi\theta_n)\eta^1_n&=& E \eta^2_n\nonumber\\
\eta^2_{n-1}+2\lambda cos(2\pi\theta_n)\eta^2_n&=& E \eta^1_n
\end{eqnarray}
We determine the critical $\lambda$ ( the threshold for the onset to magnetic transition)
as a function of $K$, the nonlinearity parameter of the two-dimensional map.\cite{Lieb}
The localization characteristics of the massless mode of the Ising model are studied  using 
an exact RG methodology\cite{KSrg}. In this approach, quasiperiodic models such as eqn (7) with
golden-mean incommensurability are decimated
to a renormalized model defined only at the Fibonacci sites. Renormalization flow describing
renormalized couplings at the Fibonacci sites provides an extremely accurate tool to distinguish
extended, localized and critical states. Trivial fixed points  of the RG describe extended states while
critical states correspond to nontrivial fixed points. As shown in Fig. 3, nontrivial $6$-cycle 
( which also corresponds to six-cycle of the wave function $\eta_n$ as shown in
Fig. 1) degenerates to trivial fixed points
at certain special parameter values.
The origin of these trivial fixed points, has been traced to
a {\it hidden dimer} in the quasiperiodic iterates describing the golden-cantorus.\cite{dimerlong}
At these points, the relativistic massless mode of the Ising model is ballistic.
Therefore, relativistic fermions may become delocalized analogous to the nonrelativistic case
due to dimer-type correlations.

The Fig. 3 shows an interesting interplay between the magnetic transition and the
ballistic transition due to dimer-type correlations: the ballistic transitions where the relativistic
mode is propagating seems to be sandwiched between two peaks corresponding to strong enhancement
of ( possibility divergent ) strength of the inhomogeneous field needed for the onset to LRO.
This phenomena again confirms the view that spin-fermion relationship may be an extremely
useful means to understand the richness and complexity underlying a variety of new phenomena in
disordered systems.
\begin{figure}
\resizebox{0.5\textwidth}{!}{%
  \includegraphics{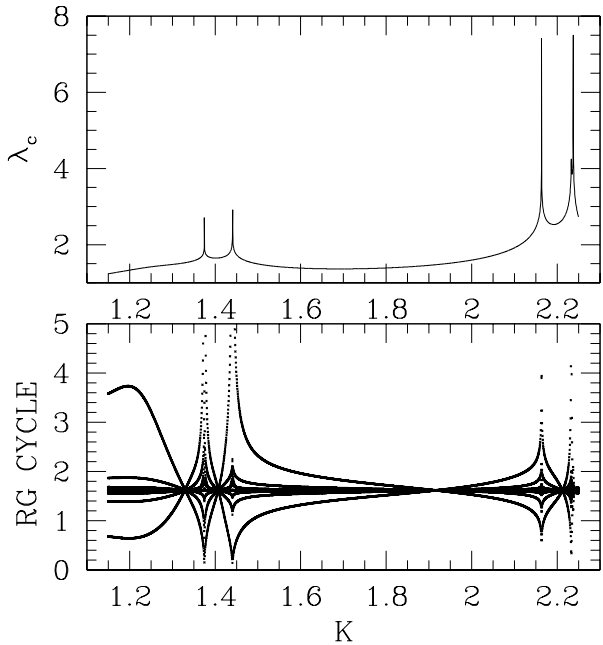}
}
\caption{(a)Critical $\lambda$ as a function of $K$ for the Ising model.
(b) RG $6$-cycle showing the variation in the renormalized coupling at Fibonacci iteration of the RG flow for $E=0$.}
\label{fig:3}
\end{figure}

One-dimensional quantum spin chain in a transverse field at the onset to LRO
describes two-dimensional layered Ising model.
Therefore, our study relates universal aspects, described by Onsager universality,
of two-dimensional Ising model to the universal 
aspects of the two-dimensional
Bloch electron problem described by the Harper equation.
This paper establishes a relationship between two important systems where geometry and integrability
are of central importance.
We believe that our results are valid for a variety of disorders including
a large class of pseudorandom as well as random cases.

Finally, it should be noted that spin-fermion correspondence is also valid for time-dependent models.
Kicked Harper model that has been extensively studied in quantum chaos literature\cite{Dima,PSS}
also describes XY spin chain in a periodically kicked inhomogeneous magnetic field.\cite{PS} 
By exploiting the spin-fermion correspondence,
various new results in the kicked Harper model can provide a new way to understand
many surprising results.\cite{SP}
One of the interesting results is the independence of the critical exponents with respect to the discommensuration
parameter $\sigma_g$ in the limit $\sigma_g \to 0$. This defines a new aspect of the usual universality
statements for the spin systems at the onset to magnetic transition and hence
broadens the concept of universality to include disorder as well as time dependence.

This research is supported by a grant from National Science Foundation,
DMR-0072813. I would like to thank Alain Comtet for bringing to my attention the problem
of disordered Dirac fermions.

\end{document}